\documentstyle[11pt,paspconf]{article}

\begin{document}

\title{Spontaneous and Induced Star Formation in the LMC}

\author{Yu.N.Efremov}  
\affil{Sternberg Astronomical Institute, MSU, Moscow 199899}

\author{B.G.Elmegreen}
\affil{IBM Watson Researcher Center, Yorktown Heights, NY 10598}

\begin{abstract}
Spontaneous  and triggered star formation in the LMC is discussed with
data on star clusters ages and positions. The  supershell LMC4  and the 
stellar arcs in the same region are  suggested to be triggered by  GRBs,
the progenitors of which might have escaped
the old elliptical cluster NGC1978,
close to which are a number of X-ray binaries and  the  SGR/SNR N49.  
\end{abstract}

to appear in IAU Symposium 190, "New Views of the Magellanic Clouds,"
ASP Conference Series, eds. Yuo-Hua Chu et al., 1999.  (a conference
held in Victoria, British Columbia, Canada, 12 - 17 July 1998)


\section{Stochastic star formation}

The LMC is the best site to study large-scale features of star formation
as displayed by mutual positions and ages of star clusters. The ages are
now available for about 600 clusters. 

Efremov and Elmegreen (1998a) demonstrated recently that the larger the
mutual distance between these clusters, the larger their age difference.
This relation may be considered part of a general correlation between the
size of a star formation region and the duration of star formation
there. For example, OB subgroups each form in $\sim3$ My, the whole OB
associations that surround them form in $\sim10$ My, and the star
complexes, such as Gould's Belt, that include the OB associations, form
in $sim30$ My; each scale is larger than the previous by a factor of 10
(e.g., 3 pc, 30 pc, and 300 pc). Thus the duration of star formation
scales approximately with the square root of the region size in this
example (see review in Elmegreen \& Efremov 1999).

Such scaling is expected for star formation in turbulent gas, because
star formation usually operates on a time scale proportional to the
cloud or clump crossing time, and this crossing time scales with
the square root of region size. Thus the formation of hierarchical young
stellar groups results from hierarchical cloud structure. There is no
preferred scale for cluster formation. 

\section{Triggered stellar arcs in the LMC4 region}

The LMC harbours what is often considered to be the classical 
region of triggered star formation, the LMC4 supershell identified
and photographed in H$\alpha$ by Meaburn (1980). This region is sometimes 
erroneously called Shapley's Constellation III, but that title should go 
to the region around the NGC 1974 cluster (see Efremov \& Elmegreen 1998b).
Three or four gigantic arcs of stars and clusters are known in the LMC4 
region  (Hodge 1967).
The most obvious arc was considered by Westerlund and Mathewson (1966): it is
inside the southern part of the supergiant HI hole/shell, found
by McGee and Milton (1966), Domg\"orgen et al. (1995), and Kim et al. (1997).  
Westerlund and Mathewson (1966) ascribed the origin of the stellar arc and HI
features to the outburst  of a Super-Supernova, following Shklovsky (1960).  
Later on, the region
was considered the best manifestation of triggered self-propagated star
formation (Dopita et al. 1985). However, there are serious difficulties 
with this interpretation if the brightest stars in the center of the
HI ring are considered to be the triggering cluster, because these stars 
are too young (Olsen et al. 1997; Braun et al. 1997).   

    Recently, Efremov and Elmegreen (1998b) suggested that two well-shaped 
arcs  in this region formed by triggered star formation in
gas that was swept-up by centralized sources of pressure.
The strictly circular shapes of both arcs are the strongest evidence for this.
Six coeval AI stars near the centre of the larger arc (called Quadrant)
were suggested to be the remnants of an association, including O-stars, which
swepted up the gas in the larger region starting $\sim30$ My ago. A small and
younger cluster near the
center of the smaller arc (Sextant) was shown to be responsible for that one.
The positions of various features in this region are shown in
Figure 1 in Efremov and Elmegreen (1998b). 

   These centralized stellar sources of pressure could produce both young 
stellar arcs at the right time and position, as Efremov and Elmegreen
(1998b) demonstrated, yet the general picture is still not satisfactory.
The main questions 
remaining are (1) why are there no giant stellar arcs or rings
around other, even more rich, clusters in the LMC,  
(2)  why are all of the stellar arcs in the LMC  close to each
other in only this place, (3) why are there just arcs and not
full stellar rings, and (4) why are the arcs in the region of
the largest and deepest HI hole in the LMC?

The recent identification of Gamma-ray burst (GRB)
afterglows in distant galaxies revived the possibility that single
super-explosions can produce large shells and
trigger star formation (Efremov, Elmegreen and Hodge 1998; Perna and
Loeb 1998). There may not even be a central cluster or evidence of 
an extragalactic cloud impact in the triggered region.  Indeed, 
the arcs in the LMC4 region, and the whole HI supershell as well, could
be produced by GRB-like explosions (Efremov, Elmegreen and Hodge, 1998).  

   This explains item (1) above, yet not the other three. 
Why would all the peculiar outbursts occur in the same region of the LMC?  
An answer to this question might come if we consider  
the common assumption that explosions of some GRBs are the result
of merging components of close binaries that include a neutron star,
black hole or  white dwarf.
This hypothesis was recently advanced  by Efremov (1998)
to account for the stellar arcs and LMC4 supershell;
the binaries  could have escaped from NGC 1978, which is a massive cluster 
of intermediate age in the same area of the LMC.

\section {NGC1978 as the origin of GRB progenitors}

The high rate of occurrence of  X-ray binaries (with one component a
neutron star) inside dense globular clusters is well known (e.g. Baylin 
1996). It was explained long ago as consequence of
the high probability of formation of close binaries after tidal captures
in the dense cluster (e.g. Shklovsky 1982, Davies 1995). It was also shown
(McMillan 1986) that a large 
number of tidally captured binaries  may escape a dense old cluster 
as the result of three-body collisions. 
Recently Hanson and Murali (1998) suggested that stellar encounters
in globular clusters were able to produce not only millisecond pulsars
but also binaries that evolve into GRBs.

The formation of a neutron star  after a SN explosion 
in a binary system leads to a high recoil velocity, the most likely value of
which is 150 - 200 km/s  (Lipunov et al. 1997). Such high velocities
would spread out any future GRB  over 
very  large distances around the paternal cluster. 
Even smaller velocities would disperse the GRB progenitors significantly, 
because the binaries may take 100 My before they merge to give a GRB 
(Lipunov et al. 1997).
We therefore suggest that the relics 
of GRB might be observed in the kiloparsec-scale regions surrounding 
old dense clusters.

NGC 1978 is a rich and old cluster, though much younger than classical
globulars in the Milky Way and the LMC.  Its age is about $2\times10^9$
(Bomans  et al. 1995), and it is the richest such cluster
in the LMC. Indeed, it has a few hundred red giants with masses of 
around 1 M$_\odot$!

    NGC 1978 is also unusual in its extremely flattened shape
(Geisler and Hodge, 1980). This may indicate a formation process involving 
the merger of two clusters, especially because no rotation has been
detected (Fisher et al. 1992).  Besides the shape, there is no other 
evidence for merging, yet Kravtsov (1998) fond some signs of abundance
differences in two parts of the cluster. 
The process of merging surely increases the probability
of stellar encounters, leading both to the formation of close binaries and 
to the escape of many stars from the resulting cluster (e.g. Oliveira et
al. 1998). The observation of a large number of blue stragglers in NGC 1978 
(Cole et al. 1997) might  also indicate a high rate of stellar encounters there.

There are also  other objects near NGC 1978  that are binary stars with 
a compact component and are therefore related to GRB progenitors if they are not
progenitors themselves. 
Three  X-ray binaries are within 20' of  NGC 1978
and more are in a wider surrounding, as is evident from 
Haberl (1998).  These objects are ascribed mostly to
high-mass binaries, yet  some classified as Be/X-ray binaries might
contain white dwarfs (see Neguerula 1998).

Moreover, the famous Gamma-ray burst of March 5, 1979 is also at about
20' to NW  of the cluster. It is now known as the Soft Gamma Ray (SGR)
repeater SGR 0526-66, and it is an X-ray binary, i.e., a stellar remnant
of a SN inside the young remnant N49 (Danner et al. 1998).
   
This is consistent with the conclusion of Nakamura (1998) that low energy GRBs 
leave behind SGR repeaters. Nakamura explained in this way  the
properties of SN1998bw (of b/c type), which was a bright Supernova 
and also a GRB with low energy (Woosley et al. 1998).
All three well-studied  SGRs are indeed connected with young SNRs
(Kouvelitou et al. 1998). The connection between SGRs and GRBs
was recently suggested by Wang and Wheeler (1998), who noted that 
it is compatible with the suggestion by Hanson and Murali (1998) on the
origin of GRB progenitors in globular clusters. 
  
 We propose that the outburst that  produced SGR 0526-66 plus SNR N49
was similar to the SN1998bw/GRB980425 event, and also that  events like
these produced the stellar arcs near LMC4.
Indeed, the Quadrant and Sextant arcs might be produced by the stellar 
winds and supernovae
from one to five dozen O-type stars (Efremov and Elmegreen 1998b), which
corresponds approximately  to 10 to 50 common SNs. 
The energy from this is comparable to or a bit below  that of a
common GRBs. An energy  10-100 times  higher was necessary
to form the entire LMC4 HI hole; this is within  the energy range
observed for GRBs.
We noted that second rich  intermediate age  cluster in the LMC, NGC1806, 
is inside the HII supershell LMC6, in the region of low density of HI.   

\section {Conclusions}
  
Stellar arcs near LMC4, and the LMC4 HI hole itself, might have been 
produced by GRBs whose progenitors originated in the intermediate-age,
nearby, globular cluster NGC 1978. 
GRBs like this may occur anywhere in a galaxy near such a dense cluster, 
whereas sequential SNs in young clusters might only occur in the spiral 
arms and near the planes of galactic disks. Indeed, the whole LMC4 region
and other similar patches of star formation with giant stellar arcs 
(as in M83 and NGC6946 -- see Efremov, Elmegreen \& Hodge 1998)
might be produced by GRBs.  The arcs might be only parts of circles (e.g., 
Hodge 1967, Efremov 1998) because the GRBs explode outside the galactic plane.  
The abundance of SNs in both M83 and NGC 6946 might be one more 
indication of a connection between certain SNs and GRBs.

One important issue that is still unresolved is why other dense 
globular clusters in the LMC and elsewhere have
no similar concentrations of GRB relicts.  For example,    
Ciardullo et al. (1990) suggested that the large number
of X-ray binaries in the bulge of M31 might be the result of ejection from the
globular clusters there.  The  unique properties  
of NGC 1978 that might have led to the observed stellar arcs are
its elliptical shape (suggesting a 
merger) and its high abundance of red giants (which are more massive stars than
in classical globulars). These properties are plausibly connected with the 
occurrence of  surrounding GRB relics and X-ray  binaries.
Perhaps the compact binaries ejected from this cluster
evolve not only to X-rays sources, but to SN/SGR and GRBs as well.

\end{document}